# Influence of the KK graviton decay into hh on the triple Higgs measurement at LHC

September 28, 2026

Alain Le Yaouanc[1], François Richard[2]

Université Paris-Saclay, CNRS/IN2P3, IJCLab, 91405 Orsay, France

***Abstract***: *Evidence for two KK graviton candidates has been previously reported at 380 GeV and 700 GeV. They would provide **graviton factories** in an e+e- collider. Following a Randall Sundrum interpretation, two extra resonances should appear at 1000 GeV and 1322 GeV. Recently ATLAS, in its search for triple Higgs coupling, has reported an excess in that mass region which comforts this prediction. Local cross sections are therefore clearly in excess of the standard predictions which implies **μHH>1**. While still marginally significant, this effect appears in a mass region with low background which allows to expect good prospects of discovery with RUN3 data. Such a result could therefore allow to confirm the existence of a series of KK graviton resonances observable at LHC. An **optimised strategy**, which includes **H(95)** is proposed to do so. Other opportunities appear in searches for heavier KK graviton recurrences decaying into ZZ/WW in semi-leptonic and fully hadronic modes. Prospects for measuring heavy KK graviton decays* $G_{KKi}$***→G380G380*** *at HL-LHC are discussed. Indirect evidences for the RS model coming from precision measurements are also presented.* ***KK lower mass limits*** *for other KK particles are deduced from precision measurements at LEP/SLC and LHC.*

## Work presented at LCWS2025 in Valencia and at IRN Terascale meeting in Orsay, April 20-22nd 2026

1 Alain Le Yaouanc <alain.le-yaouanc@ijclab.in2p3.fr>
2 François Richard <francois.richard@ijclab.in2p3.fr>

# Introduction

The present priority of the LHC program seems to be an attempt to reconstruct the SM Higgs potential. This work uses very sophisticated methods to select efficiently SM Higgs boson pairs hh. Such a measurement cannot ignore scenarios where hh can be produced by BSM resonances. In reference [1], we show some evidence for such heavy resonances, the strongest one being a sequence of KK graviton resonances, as predicted in the Randall Sundrum (RS) model with some distinctive features from this model which explains that, contrary to expectations, candidates with masses **below 1a TeV** are still not excluded.

Our best candidate for a BSM resonance has a mass of 700 GeV and is described in [1]. LHC observes **9 significant excesses of events around 700 GeV**. This resonance has a total width of ~20±10 GeV precisely measured in **$ZZ \to \ell^+\ell^- + \ell'^+\ell'^-$,** $\gamma\gamma$ **and e+e-**. The ZZ angular distribition is incompatible with a scalar interpretation, therefore suggesting a spin J=2 KK graviton interpretation of the Randall Sundrum type. The presence of an additional candidate around 380 GeV, observed in **5 significant excesses** [1], agrees as being the first of a sequence predicted by such model. These candidates should decay in about 25% of the cases in hh and therefore contribute to the Higgs potential measurement. The RS model predicts that the next KK resonances should be at 1000 GeV and 1322 GeV, with cross sections, respectively, 9 and 4 fb. These cross sections do not allow similar searches used for the 700 GeV case but fall into in a mass region where the hh channel has a low background, therefore providing a meaningful hh inclusive detection of the KK graviton resonance. It turns out that ATLAS, in its effort to measure the triple Higgs coupling, has evidence for this resonance. The SM contribution predicts a mass distribution of the hh channel, mainly located below 1 TeV even for large values of $\kappa\lambda$ as seen in figure 1b [3]. Figure 1a [2] predicts a total cross section or order 30 fb, with a differential cross section d$\sigma$/dmhh ~0.015 fb/GeV at 1 TeV, that is way below the KK graviton expectation.

(a) (b)

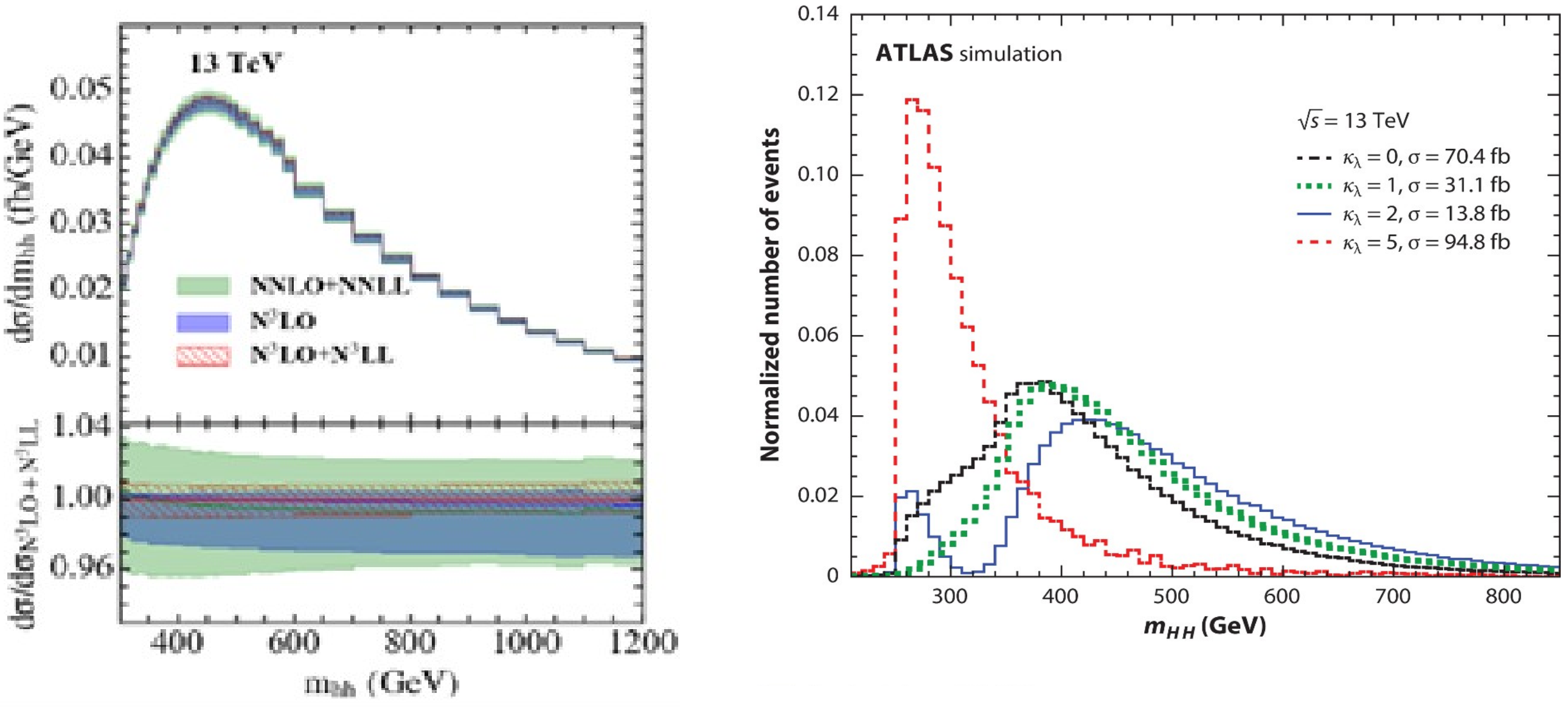

*Figure 1a: SM differential cross section for mhh distribution. Figure 1b: Same for various value of the parameter $\kappa\lambda$.*

Figure 2 from [1] shows that expected backgrounds decrease very quickly with hh masses, allowing a very meaningful measurement of hh at 1 TeV.

In the next section we will give a quantitative description on the expected mass spectrum of these KK graviton resonances and indicate how they can be measured, procuring extra evidence for the RS interpretation.

T

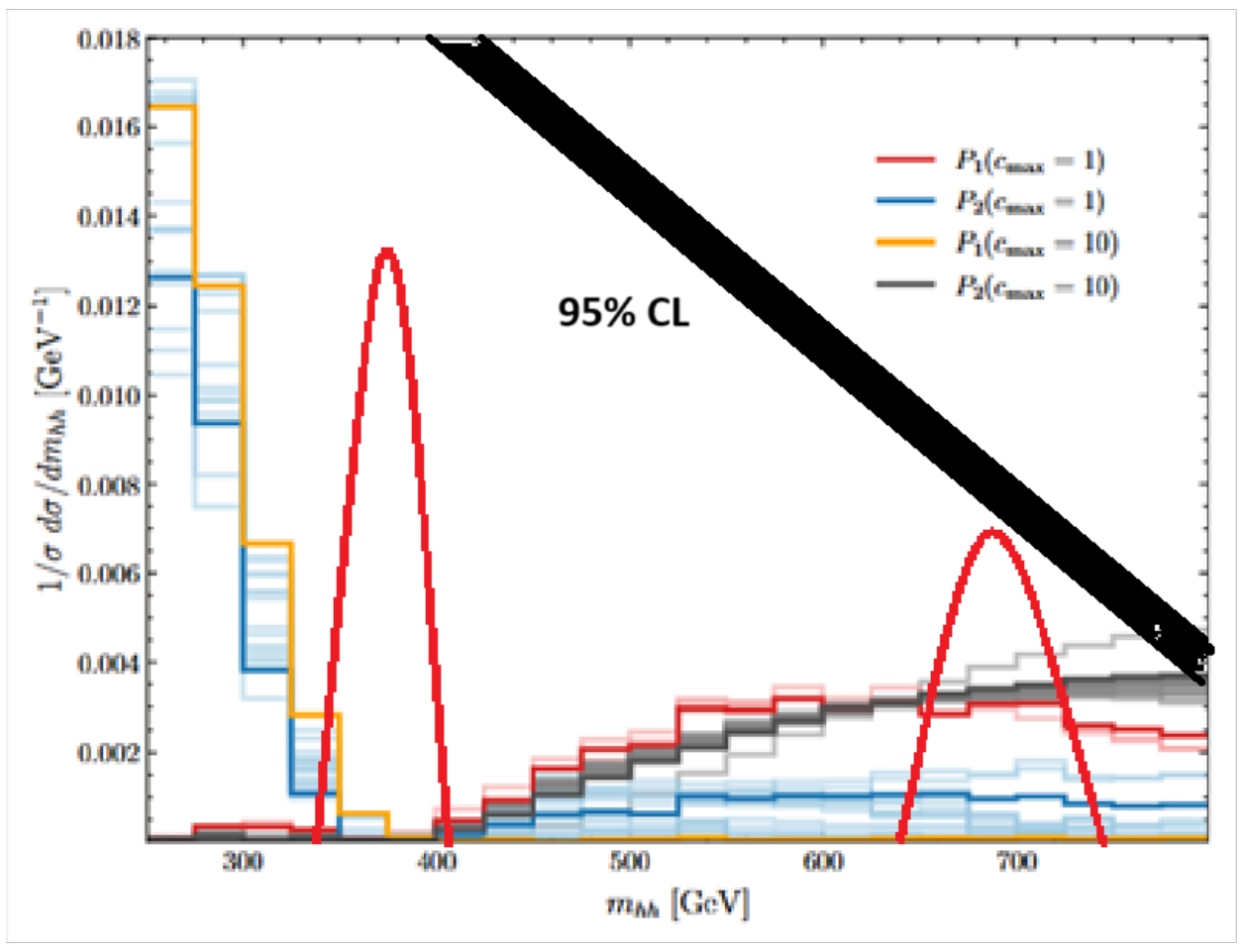

*Figure 2: Expected hh mass spectrum with various HEFT parametrisation of BSM physics with resonances above a TeV. The two red bumps correspond to expected signals due to G380 and G700. The black band corresponds to the present 95% C.L. upper limit given by ATLAS.*

# I. Predicted contributions from the graviton spectrum

(a) (b)

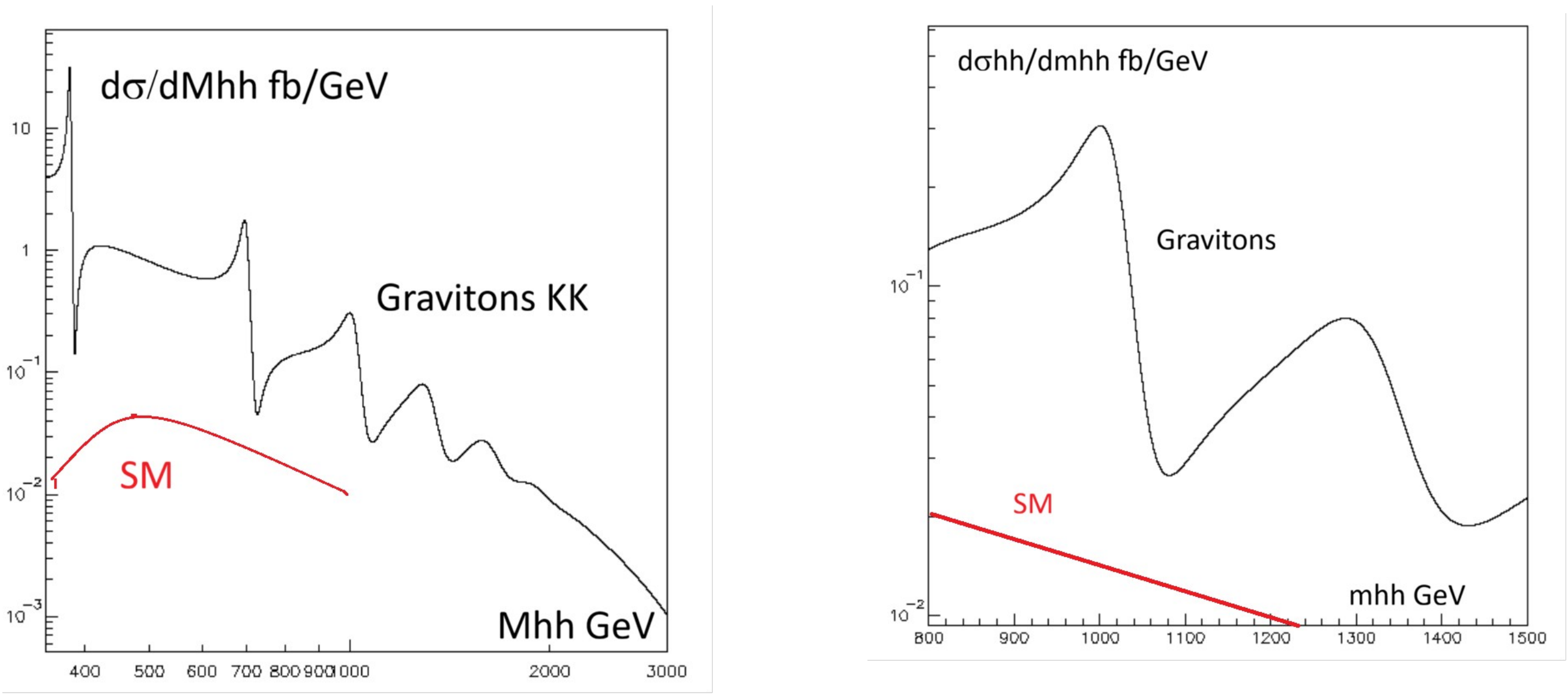

*Figure 3a: VBF differential cross section for KK gravitons. Figure 3b: VBF hh cross section for 3d and 4$^{th}$ recurrences and, in red, the predicted SM contribution.*

In the RS model, one predicts a sequence of KK graviton resonances $G_{iKK}$ which have **quantified masses** [4]. These masses are proportional to the **zeros of the Bessel function $J_1(x)$:**

**xi=3.83,7.02,10.17,13.32,16.5,19.6,22.8,25.9,29.05,32.2,35.3…**

Present indications give **mi~100$x^G_i$**,in GeV**,** while the total width itself goes like **$\Gamma$i~0.06($x^G_i$)$^3$** in GeV.

Figure 3a gives the predicted cross KK graviton hh cross sections which are assumed [1] to be dominantly VBF. The hh component has a BR of about 25 %. Figure 3b gives this contribution around 1 TeV together with the predicted standard prediction. The table below indicates the measurement methods relevant to the various mass regions.

Figure 3 predicts that resonance contributions will dominate over the SM contribution resulting in a large **μHH parameter, of order 5** which does not seem the case in spite of a clear excess, **of order 2 ,with a large uncertainty**. One can therefore suspect that the HH selection rejects most of the graviton KK resonance contribution, a mechanism which we have already observed for the ZZ analyses **where VBF→GKK→hh/ZLZL goes like $(3\cos\theta^2-1)^2$**.

**ZZ/WW e+e- $\gamma\gamma$ hh $\tau\tau$ had ZZ semi-lept VBF ZZ+WW**

| m $G_{KK}$ GeV | 380 | **700±20** | 1000 | 1322 | 1640 | 1945 | 2260 | 2590 | 2905 |
|---|---|---|---|---|---|---|---|---|---|
| $\Gamma$ $G_{KK}$ GeV | 3 | **20±7** | 60 | 130 | 260 | 440 | 680 | 1042 | 1440 |
| VBF→$G_{KK}$ fb | 360 | **100±50** | 38 | 15 | 7 | 3.4 | 1.7 | 0.9 | 0.4 |
| e+e-→ $G_{KK}$ pb | 450 | **130** | 60 | 50 | | | | | |

*Table: VBF prediction for the KK graviton cross sections and total widths. The line above indicates the relevant final states for the various mass regions.*

This table assumes that KK gravitons couple mainly to WW/ZZ/hh. The table also shows that e+e- colliders will benefit from 3 orders of magnitude improvement on the KK graviton cross sections with, however, a more limited range in mass coverage.

## I.1 Mass region below 1 TeV

This region has been the starting point for this search, giving us three major clues:

- Dominance of the VBF process over the ggF process which, as for the Higgs case, has only a top loop contribution
- The apparent dominance of the WW/ZZ modes over the top pair mode which further reduces the ggF loop contribution
- The 700 GeV resonance should have J=2 to explain its disappearance with scalar selections

From the observed effects in ZZ/WW and 2 photons final states, one can predict the whole spectrum as shown in figure 3a, in particular the relevant mass region to interpret the excess observed by ATLAS around 1 TeV shown in figure 3b. Note that there will also be non negligible contributions from mass regions below and above 1 TeV which need to be subtracted to reliably measure the parameter $\kappa\lambda$. This requires a rigorous understanding of the KK graviton spectrum which is not straightforward given that from [1] we conclude that there are important differences between our findings and the standard predictions of the standard RS model :

- Absence of direct gg and $\gamma\gamma$ coupling to these resonances as it is the case for the SM Higgs boson
- Accordingly, dominant contribution of the VBF contribution
- Suppression of the coupling to top pairs by at least an other of magnitude to agree with the observed total width, which is interpreted as a **de-localisation of the top quark (see figure 10)** with respect to EW brane to which pertain ZLZL/WLWL and hh
- This suppression reduces the rôle of top-loop contributions to ggF

Reference [31] shows that the top loop contribution is substantial in a 'top-phylic' case, of the same size as the standard value for the direct coupling. Recall that in such a case the decay in top pairs would dominate which does not seem to be the case, by almost an order of magnitude. One therefore expects that the ggF contribution should be much smaller than the standard prediction (see Appendix).

## I.2 Mass region circa 1 TeV

Concerning figure 4 from [5], two comments are of order. The channel $b\bar{b}\tau^+\tau^-$ is mainly responsible for the local excess, noting that CMS has not observed a similar effect in RUN2 data. This does not come as a surprise at this level of significance and could suggest a positive fluctuation. It is therefore too early to try a quantitative comparison with the graviton predictions.
Figure 4a suggests that, not surprisingly, the 4b final state remains in slight excess up to about 2 TeV.
One can hope that RUN3 data will allow to confirm this trend by combining ATLAS and CMS analyses.

(a)

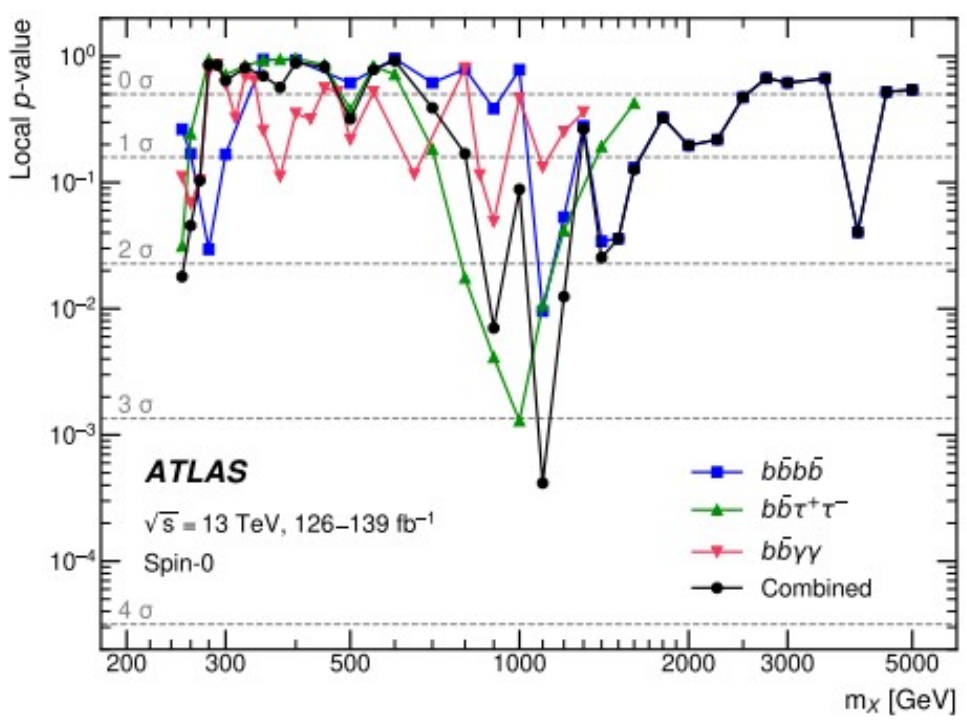

(b)

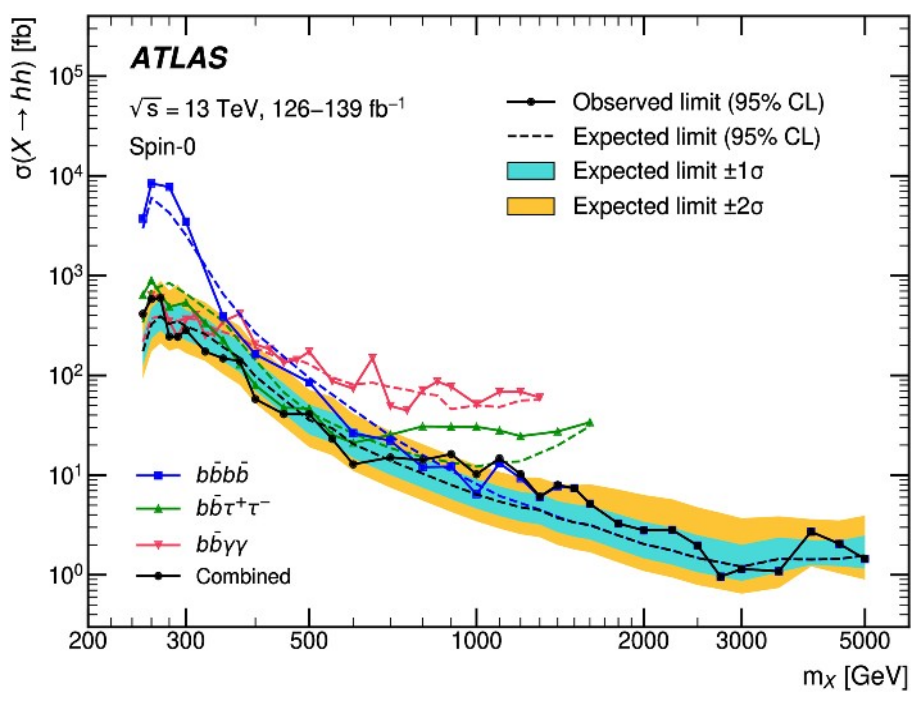

(c)

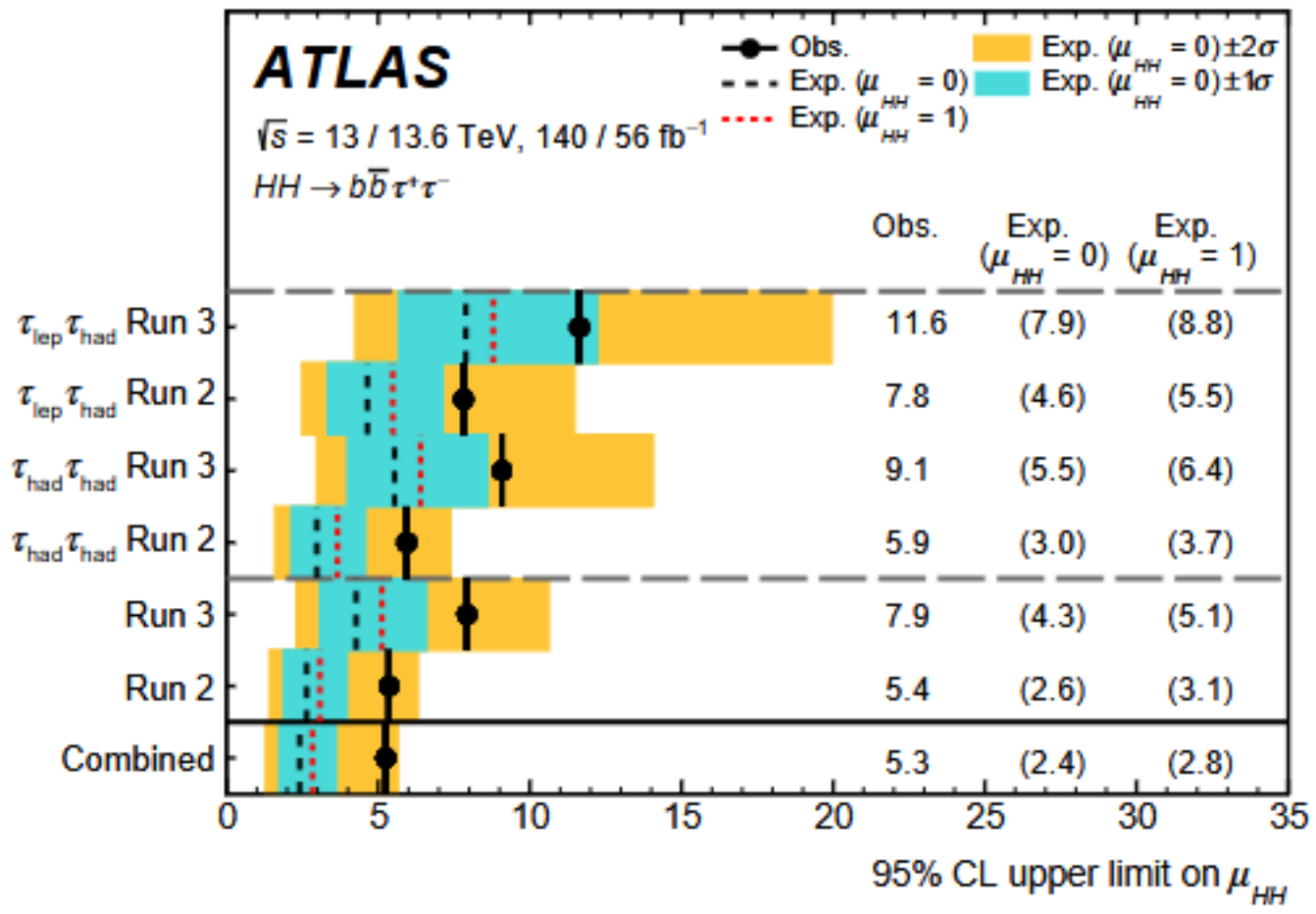

*Figure 4: Evidence for G1000→ hh. (a) indicates an excess at 1.1 TeV, mainly coming from the $b\bar{b}\tau^+\tau^-$ mode as indicated by (b). (c) shows that the excess in $b\bar{b}\tau^+\tau^-$ is confirmed by RUN3.*

CMS sees no excess with RUN2 data and a similar excess to ATLAS with RUN3 data.

CMS has reported a 2.8σ excess in $\tau^+\tau^-$ at 1.2 TeV, that is compatible with the HH excess seen by ATLAS. The measured cross section is ~3 fb which would correspond to a BR($\tau^+\tau^-$) of about 10% if one assumes a KK graviton cross section of 38 fb. This signal should also appear in the VBF mode.

Figure 5 suggests that this resonance could be wide as expected in the KK graviton hypothesis which predicts a width of 60 GeV for the 1 TeV resonance and 130 GeV for the 1.4 TeV resonance which could also contribute to this excess.

(a) (b)

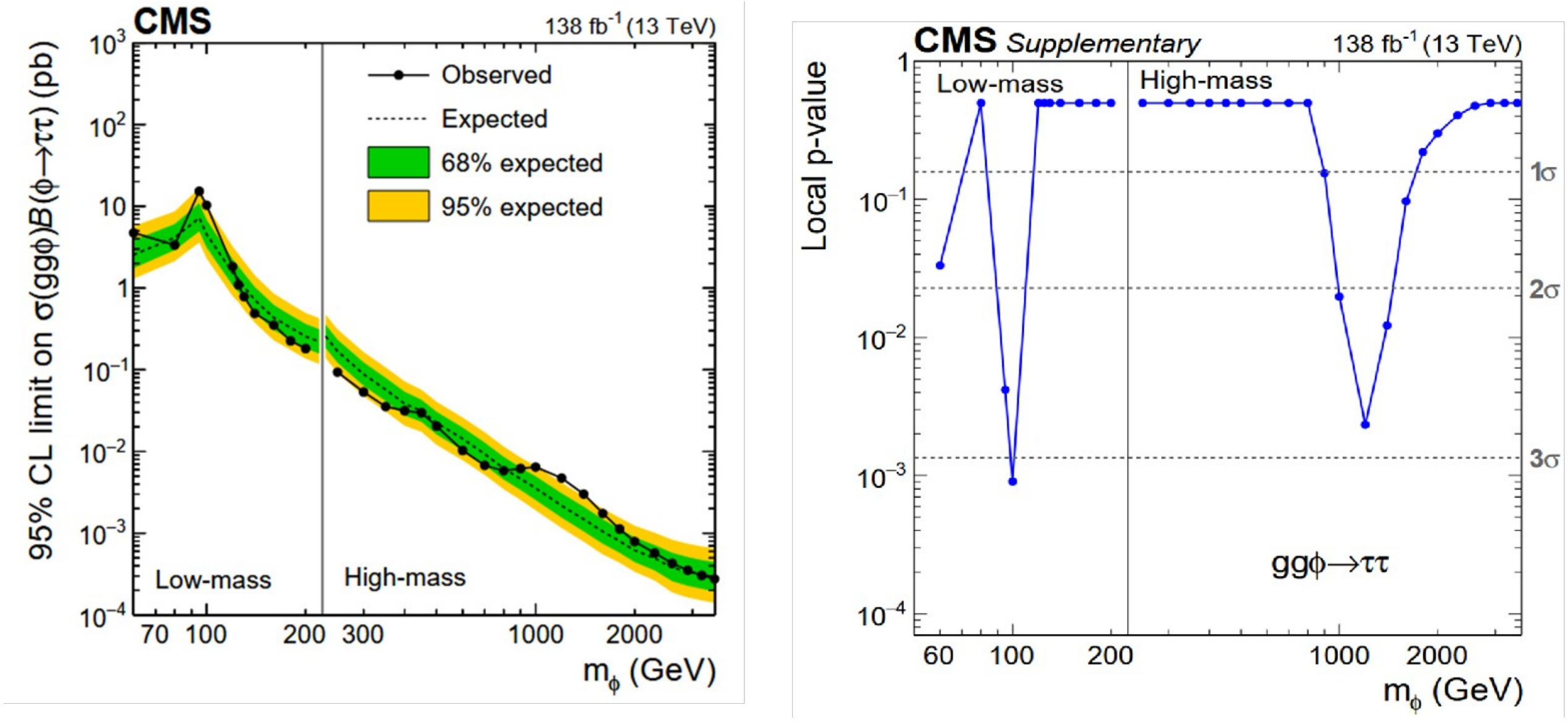

*Figure 5a from [33] shows an indication for a $\tau\tau$ contribution to the KK graviton resonance predicted at 1 TeV. 5B shows its significance.*

## I.3 Highest masses

To go well beyond 1 TeV, cross sections require using ZZ and WW modes in a more efficient way since, with reduced background, it becomes possible to relax the tight selections needed at low masses.

(a) (b)

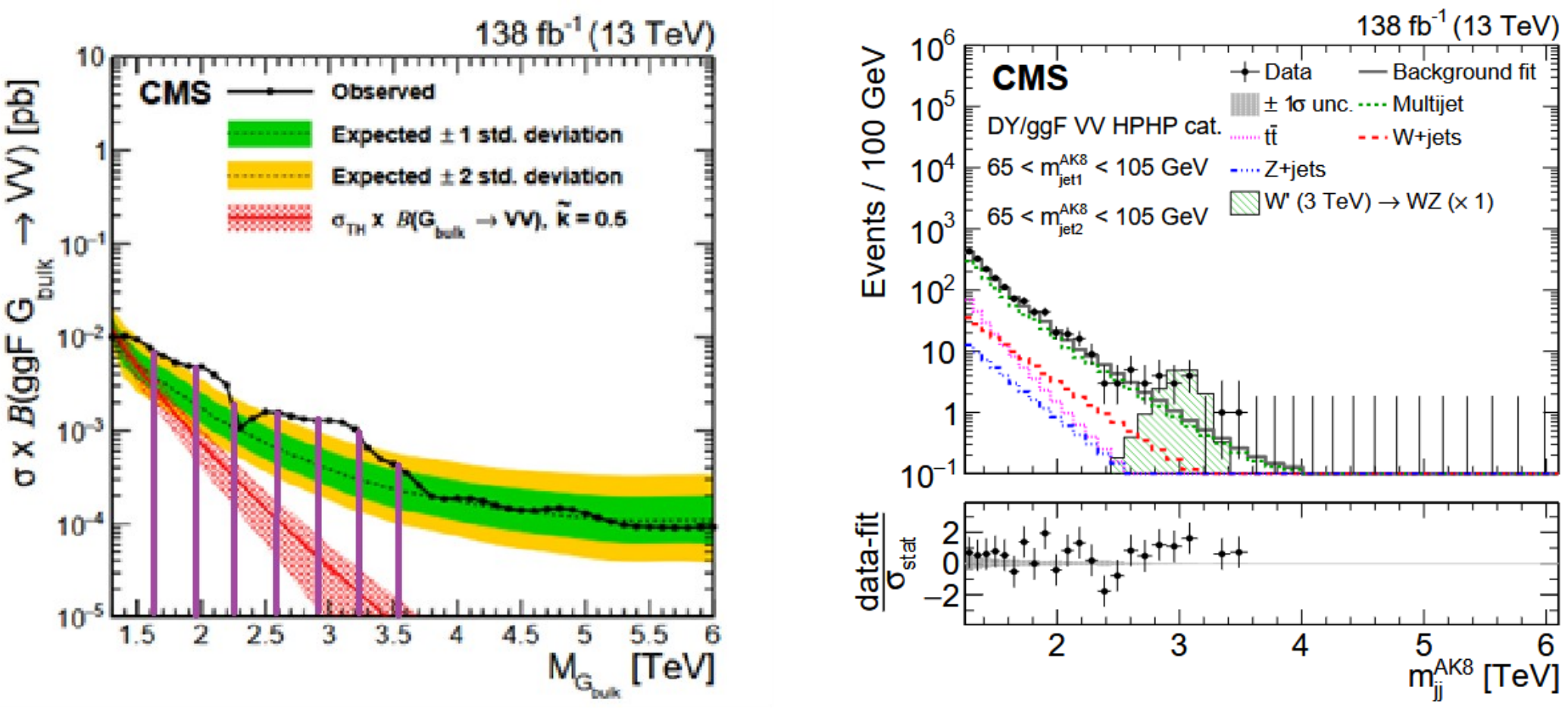

*Figure 6a: Mass limits for $G_{KK}$→WW+ZZ in jet final states. The magenta lines indicate the expected positions of the KK graviton resonances. A ~5 fb VBF cross section for masses above 2 TeV is predicted. These wide (~TeV width) resonances are expected to overlap and produce wide structure as suggested by the figure. Fig 6b: Mass distribution of the VV candidates with a visible excess around 3 TeV.*

An alternate interpretation of these indication follows from the scenario described in chapter IV where one assumes that the dominant decays of heavy KK gravitons come from the reaction **$G_{KK}$→ $G_{KK}$(380)$G_{KK}$(380)**, where the lightest KK graviton would decay hadronically through WW/ZZ/hh. It would seem that one could distinguish between the hadronic decays from WW/ZZ and those coming from G380→ZZ/WW, allowing for an enrichment of these signals with respect to the SM background, not to speak of the possible of leptonic tags which are much more frequent for these final states.

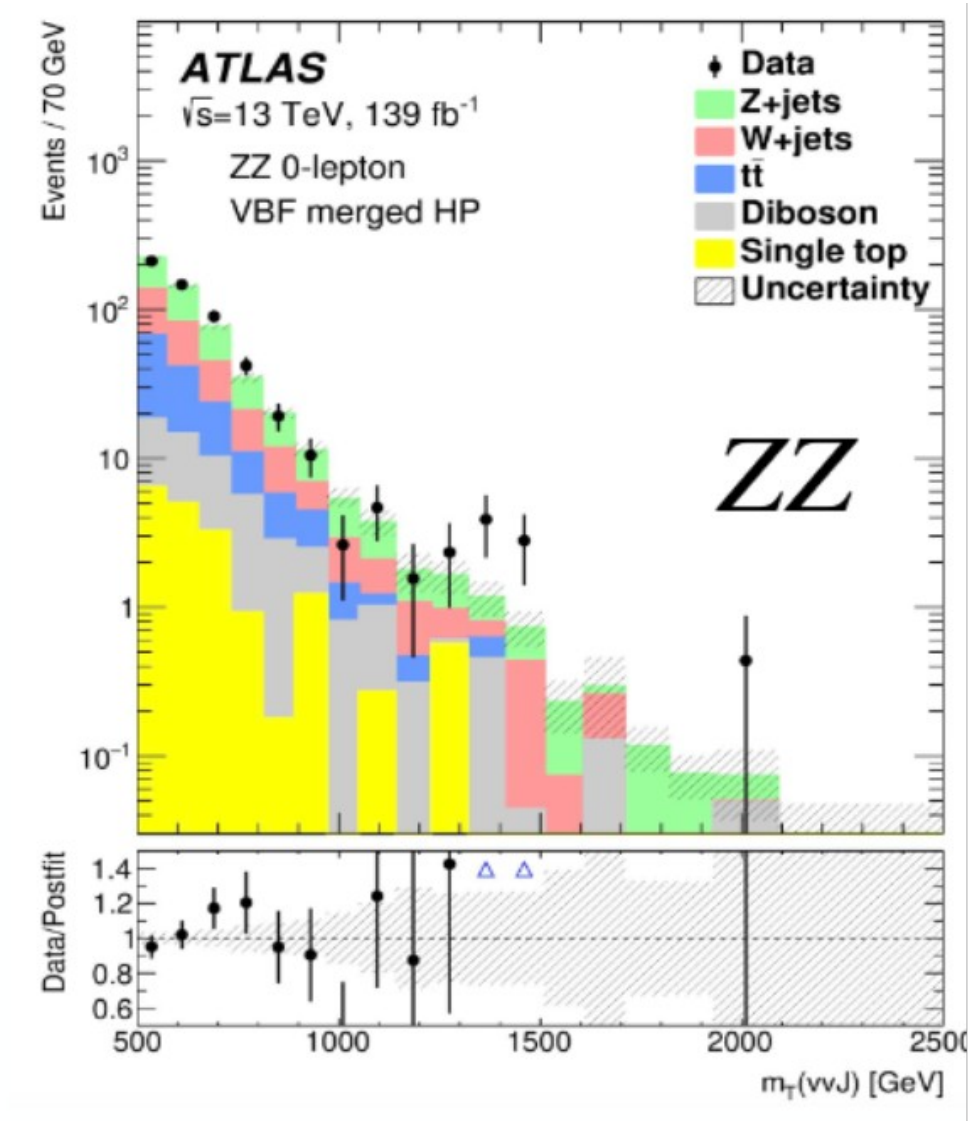

*Figure 7: VBF-> ZZ mass mass distribution using semi-leptonic modes with a HP (high purity) selection*

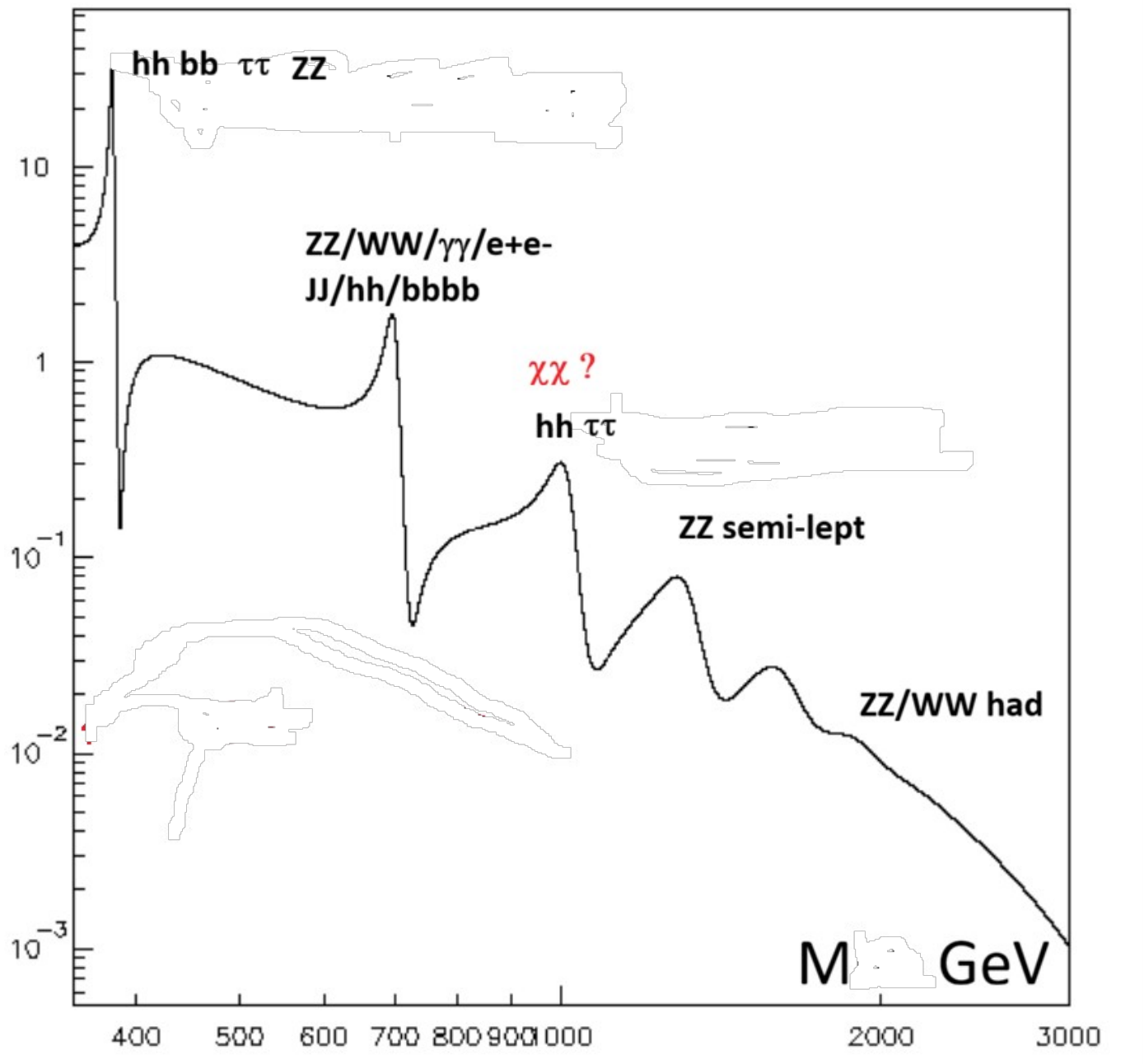

*Figure 8: Summary of the various indications from LHC together with the predicted KK graviton resonances.*

ATLAS [9] has performed a search of the **ZZ channel,** using semi-leptonic combinations. They use the **VBF process**, which is optimal according to our findings, and also observe an excess in the tail of the reconstructed mass distribution, as shown in figure 7. One can interpret the observed excess at the end of the distribution as coming from the KK gravitons with masses 1322 and 1650 GeV. The table predicts a cross section ~5.5 fb for the ZZ channels. The observed effect excess corresponds to **Aε ~2.5%**.

This low value probably reflects the fact that there are various effects (interference, reduced luminosity for the high mass tails) which reduce the effective cross section.

Figure 8 gives the mass spectrum distribution of the KK graviton resonances together with the various indications from LHC.

### I.4 A strategy for extending HH resonance searches

First of all one should take into account that VBF$\rightarrow G_{KK}\rightarrow$HH is forward peaked, which should gain in the efficiency as already pointed out.

With the light scalar H(95) observed into two photons and into $\tau\tau$ one expect a **substantial gain** by selecting $G_{KK}\rightarrow$H(95)H(95) and H(125)H(95). LHC has observed[3] GKK(650)$\rightarrow$H(125)H(90)$\rightarrow$bb$\gamma\gamma$ with 3.8 sd. These masses appear compatible with our findings in [1]. The rate for these modes will depend on the BR of H(95) which are unknown. One may however anticipate that the BR into $\gamma\gamma$ and $\tau\tau$ will be favourable, while B(bb) remains unknown[4].

In conclusion, one expects that **very substantial improvements** can be accomplished for KK graviton resonances decaying into two scalars.

# II. Precision measurements aspects

## II.1 Unitarity constraints

The reaction $W_LW_L\rightarrow W_LW_L$ receives contributions from the three SM particles Z/photon/Higgs in the s and t channels. For large s values, these contributions compensate such that the resulting amplitude does not diverge, preserving unitarity at large s. This remains true adding extra scalar doublets, given that heavy neutral scalars decouple from WW/ZZ and do not contribute at large s.

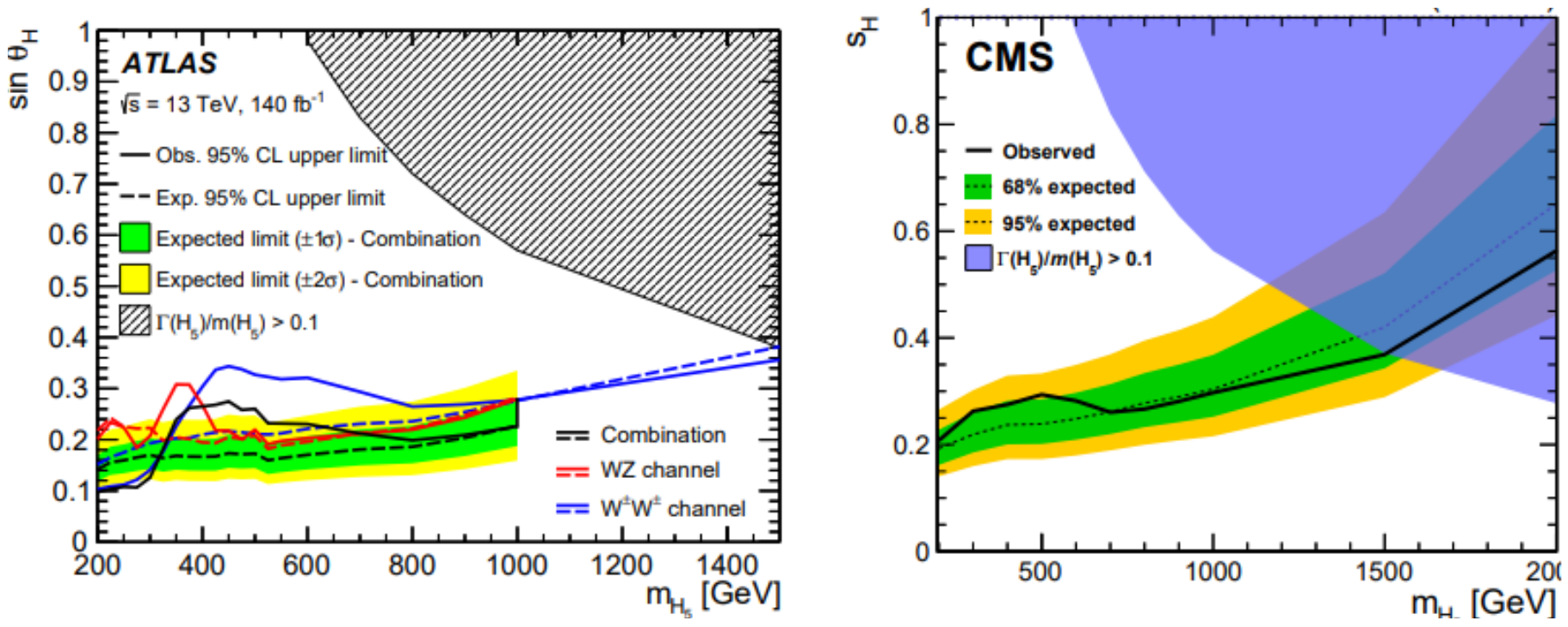

*(a)* *(b*

*Figure 9: Evidence for G+ into WZ and G++ into W+W+ in ATLAS and CMS.*

For what concerns the graviton, the lightest KK graviton resonance at 380 GeV is likely to generate an early unitarity violation [10,11] and one expects that the s channel contribution is cancelled by **u channel** contributions coupled to ZW and $W^+W^+$ resonances. Evidence for these are provided at LHC, as shown in figure 9.

The poor **mass resolution** with $W^+W^+$ does not allow a separation between the two first KK graviton resonances, hence, presumably, the wide excess (black curve of 8a) observed by both experiments [12,13].

3 See 2310.01643
4 See, for instance, 2203.13180.

This means that the **mass** of this resonance could be overestimated and in fact could be closer to the masses of G380 and $G^{+}$375, implying a smaller isospin violation. There is also an ambiguity in evaluating the $G^{++}$**cross section** given that it could integrate the two neighbouring $G_{KK}$. The cross section of $G^{++}$450 could therefore be overestimated.
Note that these states were selected as scalar resonances, following the Georgi Machacek model, while they could well be J=2 graviton recurrences, with anisotropic angular distributions, as for G700.

## II.2 Indirect effects

For this phenomenology, one defines a unique **mass scale parameter $m_{KK}$**, not directly related to the mass of given KK particles. The T and S Peskin-Takeuchi constraints imply [14] that **$m_{KK}$>30 TeV**. With **custodial symmetry,** one has **$m_{KK}$>1.8 TeV**. KK gravitons could be lighter since $m_{KK}$ is an average of KK masses. Reference [15] suggests that this can be accomplished by assuming a breakdown of universality within the kinetic terms. Neglecting these terms one would have the mass of a KK gauge particles A simply related to the KK graviton G mass: mG=1.5mA [4]. The custodial symmetry implies an extension of the SM with additional symmetries, with the **presence of a Z'** with a mass similar the KK vectors.

### II.2.1 Input from EW precision measurements

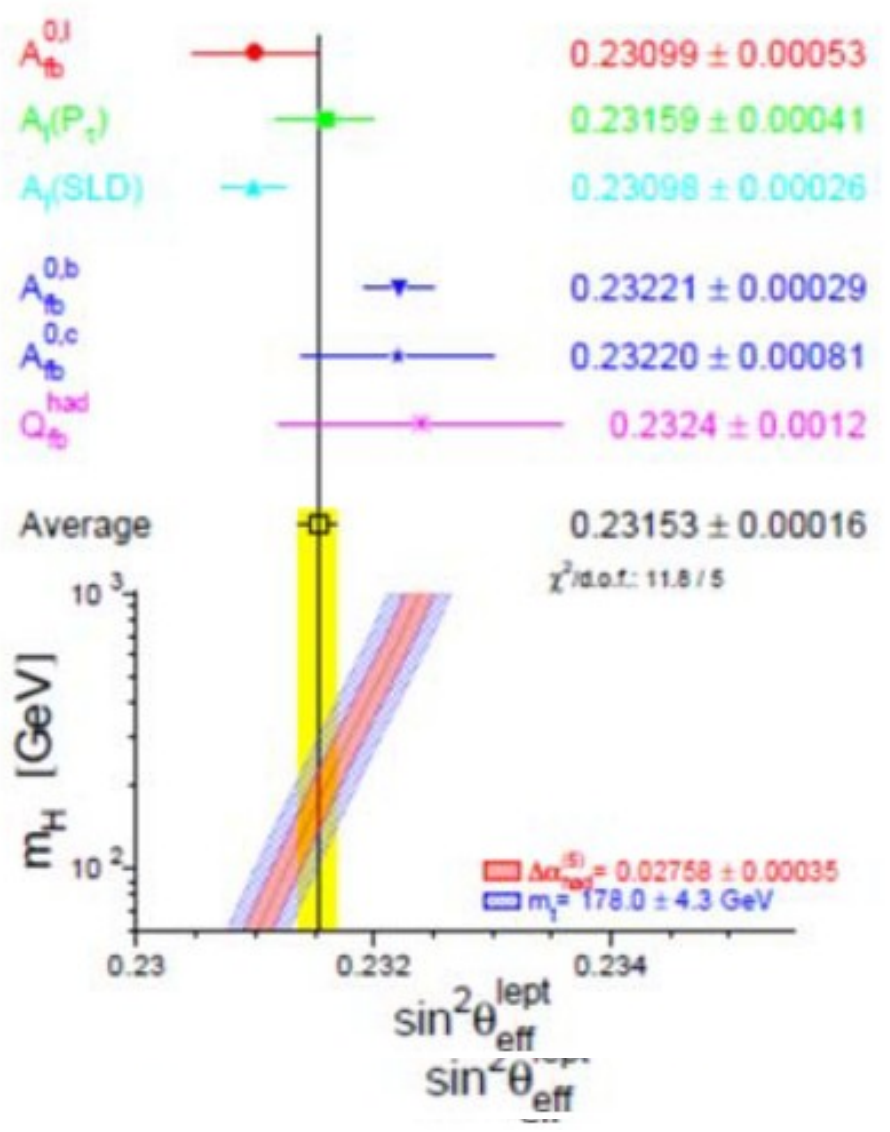

*Figure 10: LEP and SLD results on $\sin^2\theta_{lept}$*

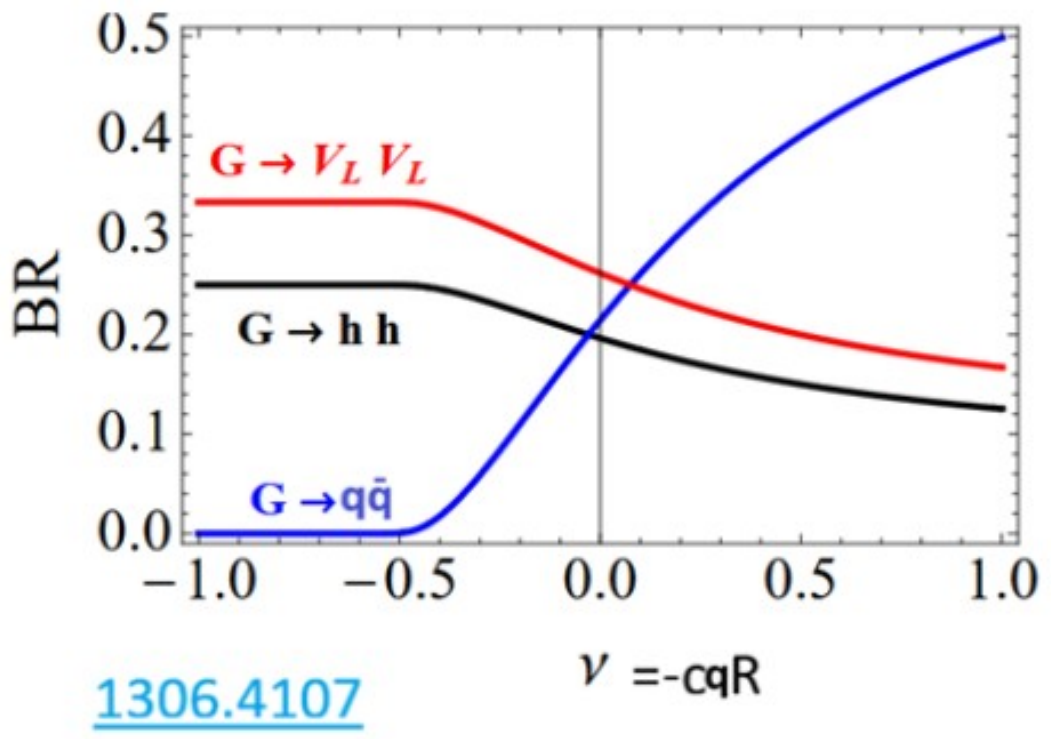

*Fig. 11: Branching ratio [17] of graviton into hh/VV and tt versus the bulk mass term ν.*

There are several deviations suggesting that RS effects could be at work: AFBb at LEP, ALR at SLD and AFBt at TeVatron. Figure 10 shows that the two most precise determinations of $\sin^2\theta_{lept}$ are away by more than 3 s.d. Averaging between these two values, one obtains an **apparent 'conformity'** with SM predictions. Interpretations in terms of RS have been provided in [16]. They can be reconciled with $Z'/Z_{KK}$ masses of order 3 TeV assuming $\sin\theta'=0.1$ for the mixing angle related to a Z'. Then one has $c_{eR}=0.56$, meaning that the coupling of $G_{KK}$ to e+e- will be vanishingly small which does not seem to be the case [1].

One can assume that $Z'/Z_{KK}$ are much heavier. Then $c_{eR}$ decreases and the coupling of G(700) to e+e- can become substantial. The dependence of the coupling to fermions [13] can be deduced from **figure 11** given that $c_{eR}=-\nu e$. Satisfying the $A_{LR}$ result of figure 10 and the indicated value of BR(ee) of G(700), requires that $c_{eR}=0.45$ and **$mZ_{KK}$=2.5TeV/sinθ'**. Since perturbativity requires that $\sin\theta'>0.1$ (meaning **$mZ_{KK}$<25 TeV**), this particle should be observed either indirectly [18] at a LC and/or directly at LHC for larger values of $\sin\theta'$. This interpretation implies that KK vectors are much heavier than the KK gravitons indicated by LHC. Note again that LEP PM have no direct implication on the $G_{KK}$ sector since one can assume that masses of KK vectors and tensors are independent.

Future Z factories, linear or circular, will allow to improve the LEP/SLC accuracies by an order of magnitude, thus fully confirming/discarding the effect seen in figure 10. If confirmed, this prediction would constitute an impressive illustration of the **mass reach of precision measurements in e+e-,** far more spectacular than the prediction for the top quark and the Higgs boson masses.

Direct searches for $Z_{KK}$ will cover the 3-5 TeV region where this resonance is expected to be narrow and therefore observable in e+e-. The next rank KK resonance is unlikely to be observed in the leptonic mode at LHC but could be observable in top pairs. The Z' particle, assumed at the same mass, is expected to be wide and only produced through bbF therefore difficult to observe at LHC [32].

## II.2.2 Effect of $W_{KK}$ and $q_{KK}$ on SM Higgs BR

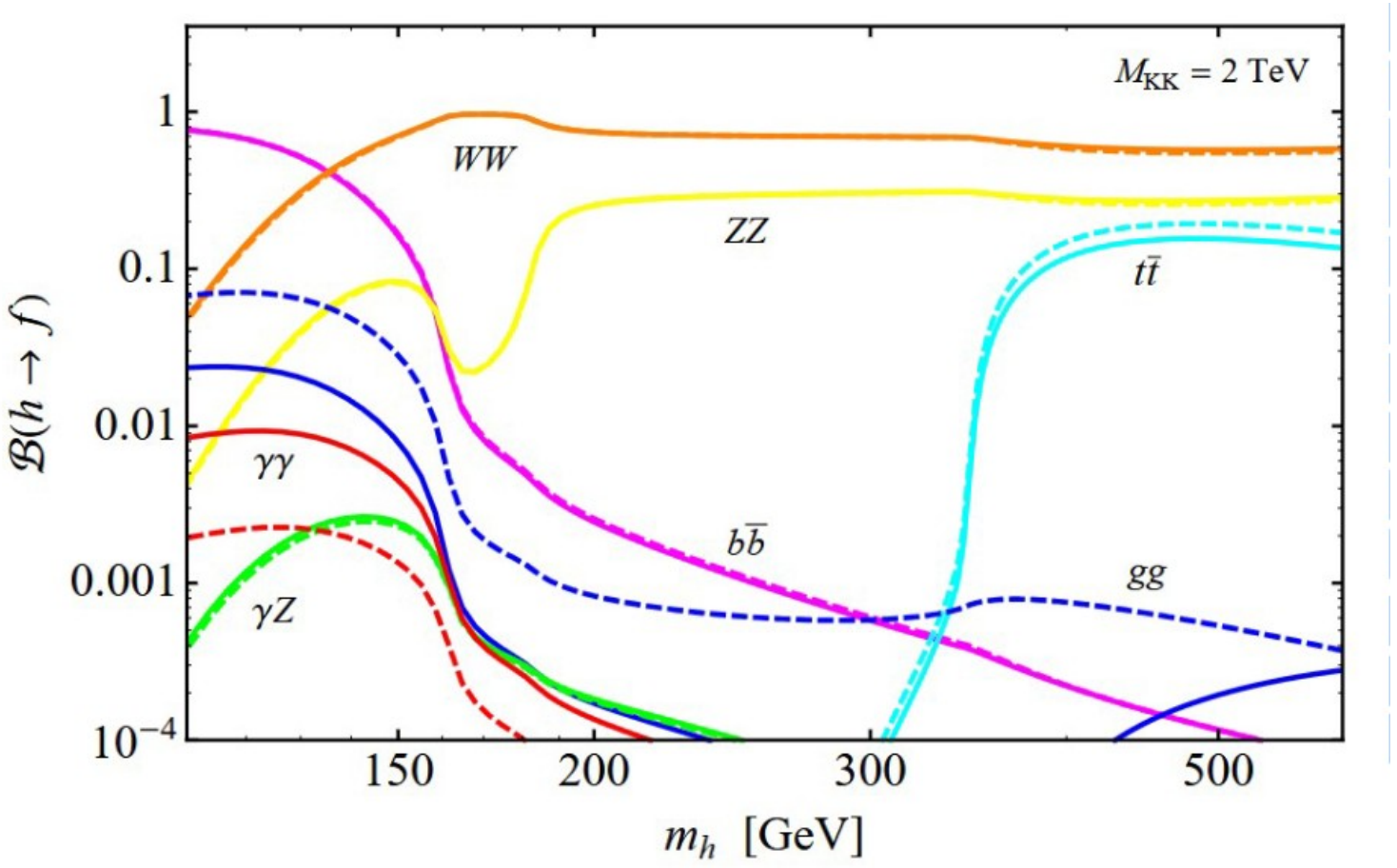

*Figure 12: Predicted BR for the SM Higgs boson [14] assuming the presence of 'light' (~2 TeV) KK particles. The dashed curves correspond to the SM and show clear discrepancies with the full curves for gg and $\gamma\gamma$ final states.*

Figure 12 shows that for a 'KK scale' of 2TeV, one expects large deviations with respect to the SM for the gg and $\gamma\gamma$ branching ratios of H125, which are excluded by LHC measurements. Processes contributing to such deviations are:

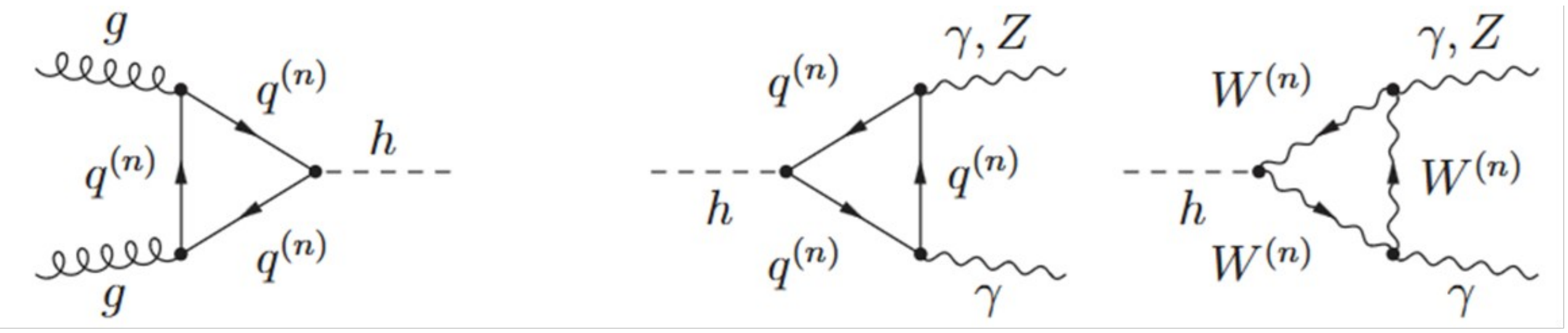

Recall however that the KK scale and **actual masses of the KK particles** are two different things. For instance the KK gravitons could be much lighter than 2 TeV. To cure the effect observed for the **gg branching ratio** it is sufficient to assume that the **KK quarks have masses above 6 TeV** while figure 12 is drawn assuming that the mass of these particles is 2 TeV.

For what concerns the **γγ branching ratio**, taking into account the gg BR constraints, one can set a lower bound on the **KK W/Z mass of order 3 TeV,** which matches with the deviation observed at LEP/SLC./Z and is reachable by LHC direct searches.

# III. Possible consequences for DM searches

Recently an analysis using **Fermi-LAT** data has provided some evidence for photon excesses which can be attributed to annihilations of **DM particles with masses circa 500 GeV.** Reference [19] has questioned the compatibility of these excesses coming from the halo of our galaxy with the limits given by dwarf galaxies. To reconcile these observations, one would need a **resonance at about 1 TeV** which couples to these DM particles. It turns out that one of the consequences of our work is that it is not unlikely to find a KK graviton resonance at the right mass which could do the job.

The process: **$\chi500\chi500 \rightarrow G_{KK}(1000) \rightarrow ZZ/WW/hh$**

could explain the origin of the excess of photons around 20 GeV observed by Fermi-LAT given that fragmentation of the hadronic decays of ZZ/WW/hh reproduces the observed energy distributions. Recall also that this resonance could decay into two photons, giving a mass peak around 500 GeV which, however, has not been clearly observed [21]. This is expected given the low branching ratio observed for **$G_{KK}(700) \rightarrow \gamma\gamma$**, at the % level, as shown in figure 1 from [22].

**At LHC** the Fermi-LAT DM candidate could originate from **KK gravitons with masses above 1 TeV**, noting that these particles are produced with a **cross section of 140 fb**. Reference [83] describes possible detection methods which select VBF events with missing transverse energy.

One expects that a **TeV e+e- collider** producing this resonance would be an ideal set up to observe these particles by measuring the **invisible width. This contribution could be measured by operating above the T1322 resonance, selecting radiative return to tag events with invisible energy.** In a SUSY interpretation, one could also observe the associated charginos which would be pair produced with a substantial cross section. Even in a compressed type scenario where the chargino and neutralino masses are very close (see for instance [20]), an e+e- machine will be able to measure this channel. It is not excluded that HL-LHC could observe it.

# IV. Searching for $G_{KKi} \rightarrow G_{KK1}G_{KK1}$ at LHC

| m $G_{KK}$ GeV | 1322 | 1640 | 1945 | 2260 | 2590 | 2900 |
|---|---|---|---|---|---|---|
| VBF fb | 15 | 7 | 3.4 | 1.7 | 0.9 | 0.4 |
| Total width GeV | 130 | 300 | 550 | 1460 | 2900 | 5600 |
| BR(G380G380) % | 2 | 12 | 20 | 53 | 63 | 75 |
| $G_{KKi} \rightarrow$ G380380 fb | 0.3 | 0.8 | 0.7 | 0.9 | 0.6 | 0.3 |

Such a process, computed in [23] and [24], would provide the ultimate proof of the existence of heavy KK gravitons and constitute an important test of the RS model. One predicts the cross sections and total widths given in above table.

This table shows that there are significant but small contributions of this mechanism which does not change our general conclusions.

Predicted decay modes are WWWW 25%, ZZWW 25%, hhWW 25%, hhZZ 12.5% and hhhh+ZZZZ 12.5%. A detailed study has to be performed. One should try to reconstruct WWWW events in purely hadronic (20%) and semi-leptonic (80%) modes which should provide a high detection efficiency and sufficient purity given the high masses involved.

To our knowledge the SM background for this topology has not been, so far, computed. One expects a contribution from the DY process:

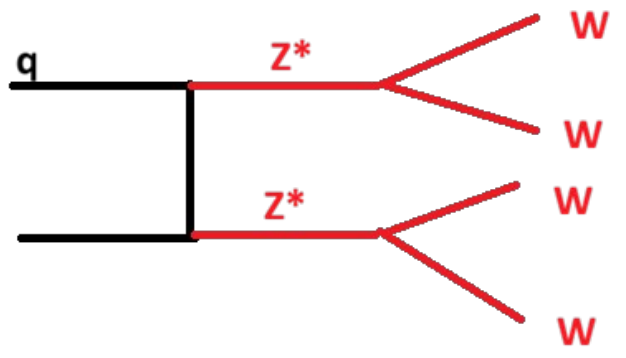

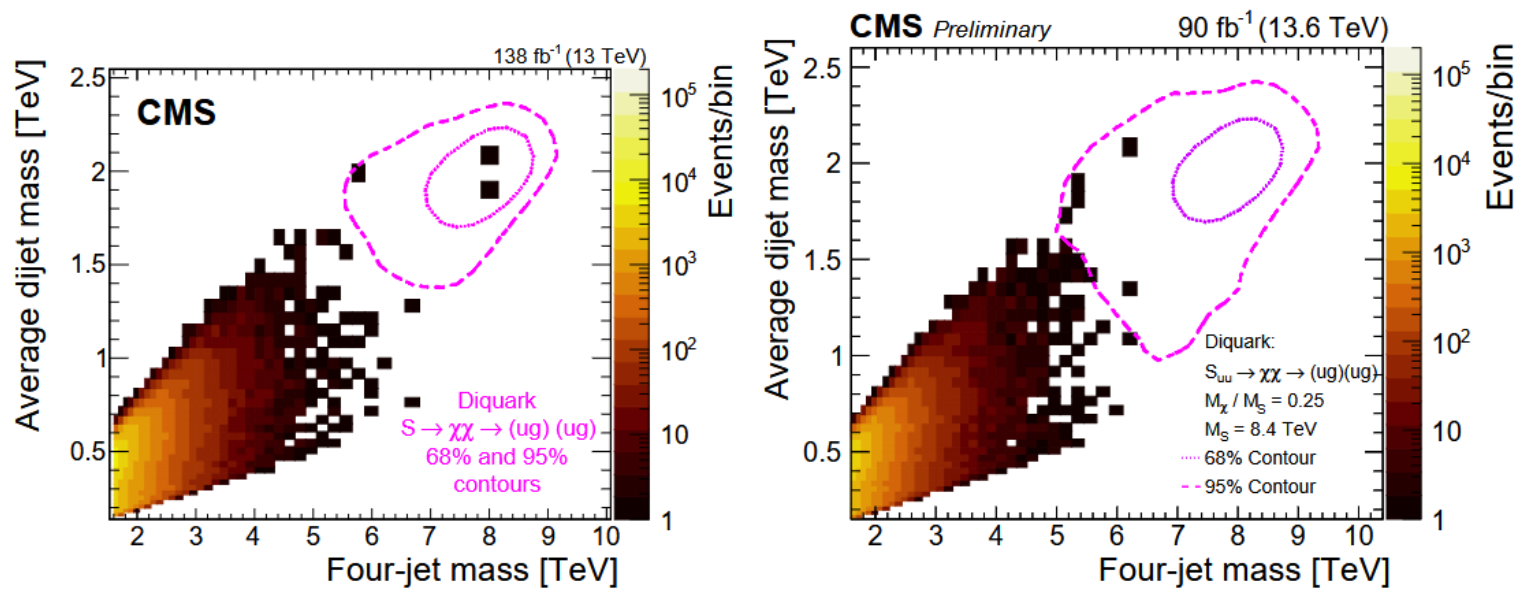

*Figure 13 : Search for 4 jet events at 13 TeV and 13.6 TeV.*

One can speculate that for KK gluons there could be a similar situation for the process with KK gluons: $q\bar{q} \rightarrow g_{KKi} \rightarrow g_{KK1}g_{KK1}$

This could explain the observation of CMS [25] reported in figure 13. Admittedly **KK-parity conservation** is invoked to discard this coupling but this may not apply to the RS case [28].

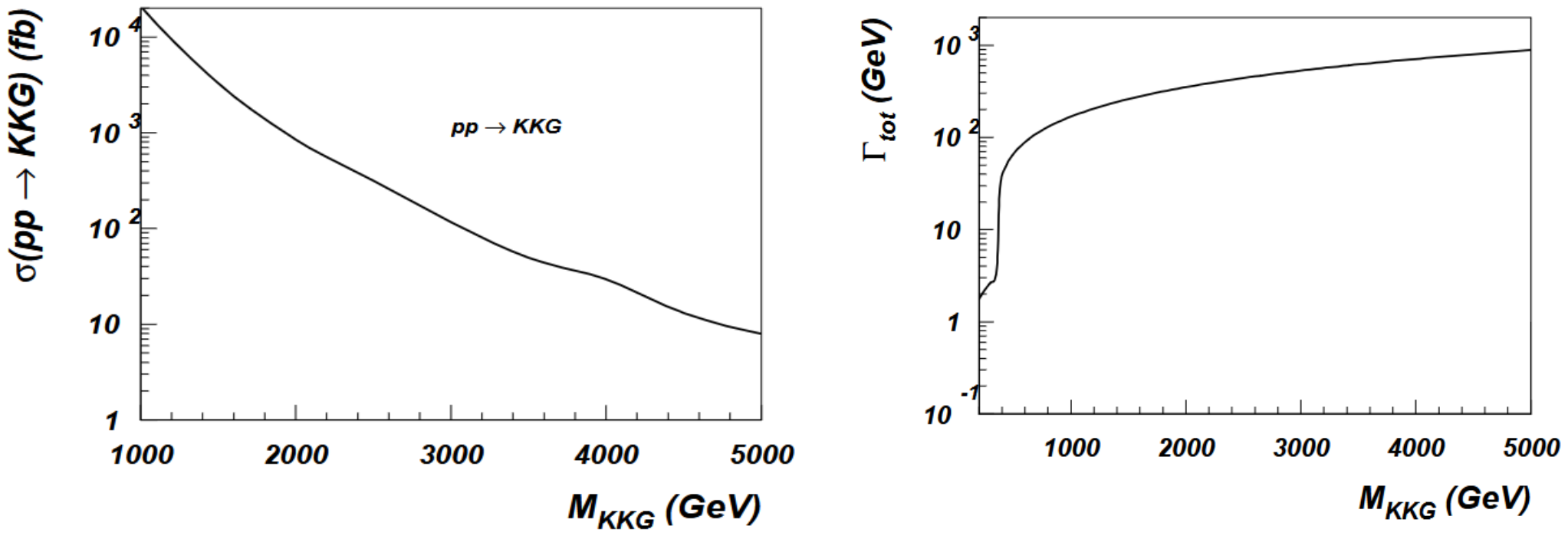

*Figure 14: Predicted cross sections and widths for a KK gluon in the RS model at LHC.*

This would mean that the lightest KK gluon weights about **2.1 TeV** which does not seem to contradict present limits. In RS, one expects that the **8.4 TeV** resonance much wider than predicted in [25]. The 2.1 TeV resonance has a mass which falls close to the predicted vectorial resonance $Z_{KK}$. There could be a second signal starting to build up around **7 TeV**. In a warped theory one would expect a sequence related to the zeros of a Bessel function $J_0$ [4]: 2.45, 5.57, 8.7, 11.84. This translates into the masses: 2.1, 4.8, 7.5, 10 TeV. These resonances should be wide and **truncated** on the higher end by the decreasing luminosity, meaning that they could be consistent with observations.

From [26] one expects a production cross section for $g_{KK}$(8.4 TeV) of a fraction of fb which is sufficient to interpret the number of observed events, given that the dominant decay mode has a good acceptance. The width of this resonance at 2.1 TeV is ~350 GeV. By analogy with the graviton case, one expects that above 4 TeV, the width should be much larger than predicted by [26].

We are of course aware of alternate possible interpretations in terms of SUSY with R parity violation [28]. A KK graviton decaying into a gluon and a KK gluon or in a pair of radions [27] is an other possibility but seems unable to provide an adequate rate if produced by VBF.

# Conclusion

There seems to be growing evidence for the presence of a string of KK graviton resonances at LHC. One can summarise the situation by saying that these resonances behave as a '**KK graviton snake**', emerging periodically in various modes which can be observed thanks to the ingenuity of the various searches at LHC. The two lowest recurrences, with masses below a TeV, are narrow and can be observed in the cleanest decay modes, in particular the ZZ mode into four leptons. Additional modes allow to provide a very solid basis for the second rank resonance G(700).

Searches performed for hh final states allow a first observation of the third recurrence G(1000). These searches look very promising but need to be adjusted to take into account the **angular distribution of a J=2 particle**. They can be eventually extended by including the **h(95) resonance**.

Higher rank KK resonances are wide and overlapping but the decreasing SM backgrounds allow to observe semi-leptonic and even hadronic final states till more than 3 TeV.

We are fully aware that experimental observations significantly differ from the RS orthodoxy but this feature can be simply reconciled by invoking **duality arguments**, assuming that, as for a composite Higgs, the preons constituting the KK gravitons are charge and colour neutral and therefore do not couple directly to photons nor gluons, which dramatically reduces the visibility of light KK gravitons at LHC.

We hope that RUN3 data will provide enough data to confirm these strong indications.

More ambitious avenues can be explored, as the discovery of other KK particles, as the possibility to measure the decay of a **heavy KK graviton** into two lower mass gravitons, G(2260)→G(380)G(380). By analogy, an interpretation of an indication of an 8.4 TeV resonance observed in 4 jets in terms of a **heavy KK gluon** decaying into pair of first generation KK gluons with masses ~2.1 TeV is proposed.

Heavy KK gravitons could play an important role in the DM sector.

**Powerful predictions** on KK Z/W (>3 TeV) and quark (>6 TeV) masses are deduced from precision measurements at LEP/SLC and LHC.

**Acknowledgements:** *We are grateful to **Yves Sirois** for providing useful infos on the LHC BSM Working Group. We have greatly benefitted from inspiring discussions with **Marc Escalier** on HH analyses.*

# APPENDIX

## Top-loop contribution to ggF->$G_{KK}$

It seems that direct couplings of gluons and photons to gravitons are suppressed with respect to RS standard expectations. One cannot however ignore loop processes which can contribute as discussed in [31]. This reference assumes that top-loops dominate this process which may not be the case for the two photon coupling which could receive its major contribution from W loops as is the case for the Higgs boson into two photons.

One can write:

$$c_{gg}^{\rm eff} = c_{gg}(m_G) + \frac{c_{tt}}{3} A_G\left(\frac{4m_t^2}{m_G^2}\right) ,$$

$$\begin{aligned} A_G(\tau) = -\frac{1}{12}\Big[ & -\frac{9}{4}\tau(\tau+2)[2\tanh^{-1}(\sqrt{1-\tau}) - i\pi]^2 \\ & +3(5\tau+4)\sqrt{1-\tau}[2\tanh^{-1}(\sqrt{1-\tau}) - i\pi] \\ & -39\tau + 12\ln\tau - 35 - 12\ln 4\Big] , \end{aligned}$$

where $\tau=4m_t^2/m^2_{GKK}$.

The first term is of order 1 if there were direct coupling to gluons, while the second term corresponds to the top loop contribution and can also reach a similat value if the top quark is maximally coupled to KK gravitons. The later does not seem the case since it would imply a total width for the 700 GeV well beyond observation. One expects instead that $c_{tt}$~0.3 and $c_{gg}$~0.

For what concerns the photons one expects:

$$c_{\gamma\gamma}^{\rm eff} = c_{\gamma\gamma}(m_G) + \frac{2}{3} Q_t^2 N_c c_{tt} A_G\left(\frac{4m_t^2}{m_G^2}\right) ,$$

The two photon contribution is too small with respect to observation, which could mean that it is unjustified to neglect the WW loop contribution which could be dominant for $c_{\gamma\gamma}$, as for the Higgs case. This approximation was perhaps only justified while assuming a large coupling of the KK graviton to top quarks. Note that ggF for G700 has a substantial imaginary part which implies deviations from a simple BW behaviour as reported in [1]. This could perhaps undermine our conclusion about an absence of ggF coupling for this resonance.

| $M_{GKK}$ GeV | $c_{gg}$ for $c_{tt}$=1 | $c_{\gamma\gamma}$ for $c_{tt}$=1 and WW=0 |
|---|---|---|
| 380 GeV | 0.9+i0.05π | 0.5-i0.05π |
| 700 GeV | -0.3-i0.1π | -0.24-i0.09π |

# Process $G_{KKi} \rightarrow G_{KK1}G_{KK1}$

Following ([23,24], one writes :

$R=\Gamma G_{KKi}G_{KK1}G_{KK1}/\Gamma G_{KKi}WW\sim\chi^2_{i11}(1-4/r^2)^{5/2}(780r^2+616r^4+52r^6+180+r^8)/936$ with $r=mG_{KKi}/mG_{KK1}$

$$\chi_{ijk} \equiv \frac{-2}{J_0(\gamma_i)J_0(\gamma_j)J_0(\gamma_k)} \int_0^1 du\, u^3 J_2(\gamma_i u)J_2(\gamma_j u)J_2(\gamma_k u)\,, \text{ where } \gamma_n \text{ denotes the } n\text{th zero of } J_1(x).$$

Above formula shows a strong dependence in r, power 8, meaning that heavy KK gravitons could significantly decay into the lightest pair G380G380.

In our case one finds the following values:

$\chi_{411}$=-56.5, $\chi_{511}$=62., $\chi_{611}$=-69.5, $\chi_{711}$=74., $\chi_{811}$=-78., $\chi_{911}$= 84. .

Great care must be taken to compute the Bessel function function $J_2(x)$ which enters in above integral. We have used the precise approximations given in [30].